# Spin transport enhancement by controlling the Ag growth in lateral spin valves

Miren Isasa[1,*], Estitxu Villamor[1], Lorenzo Fallarino[1], Olatz Idigoras[1], Anna K. Suszka[1,2], Christopher Tollan[1], Andreas Berger[1], Luis E. Hueso[1,3] and Fèlix Casanova[1,3]

[1]CIC nanoGUNE, 20018 Donostia-San Sebastian, Basque Country, Spain.
[2]current address: Laboratory for Mesoscopic Systems, Department of Materials, ETH Zurich, 8093 Zurich, Switzerland, Laboratory for Micro- and Nanotechnology, Paul Scherrer Institute, 5232 Villigen PSI, Switzerland.
[3]IKERBASQUE, Basque Foundation for Science, 48011 Bilbao, Basque Country, Spain.

*E-mail: m.isasa@nanogune.eu

## Abstract

The role of the growth conditions in the spin transport properties of silver (Ag) have been studied by using lateral spin valve structures. By changing the deposition conditions of Ag from polycrystalline to epitaxial growth, we have observed a considerable enhancement of the spin diffusion length, from $\lambda_{Ag} = 449 \pm 30$ to $823 \pm 59$ nm. This study shows that diminishing the grain boundary contribution to the spin relaxation mechanism is an effective way to improve the spin diffusion length in metallic nanostructures.

A new generation of *spintronic* devices, which rely only on the electron spin degree of freedom, are envisioned towards a future integration of logics and memory [1]. Creation, transport and detection of a pure spin current, *i.e.*, a flow of spin angular momentum without being accompanied by a charge current, are thus essential ingredients for a successful device. Lateral spin valves (LSVs) are basic *spintronic* devices that offer an attractive means to study the spin transport as well as the spin injection properties in different materials. After the pioneering studies, first by Johnson and Silsbee [2,3] and more recently by Jedema *et al.* [4,5], a large number of spin injection experiments have been reported in metals [6-22], semiconductors [23,24] or carbon-based materials [25,26]. LSVs consist of two ferromagnetic (FM) electrodes, used to inject and detect pure spin currents, bridged by a non-magnetic (NM) channel, which transports the injected spin current (see Fig. 1(a)). For the optimum performance of a LSV, it is crucial to choose a NM material in which the spin information can travel over long distances, *i.e.* with long spin diffusion length $\lambda_{NM}$, with Cu [4-12], Al [2,5,9,13,14] or Ag [15-22] being the most commonly selected metals. In order to enhance $\lambda_{NM}$, it is crucial to understand which are the spin relaxation processes that lead to the loss of spin information. It is known that, in NM metals, the spin relaxation is governed by the Elliott-Yafet (EY) mechanism [27,28], with phonons, grain boundaries, impurities or the surface being common sources for the associated spin-flip scattering [5,7,12,18,19]. A proper control of these contributions could thus help obtaining longer $\lambda_{NM}$ values.

In this work, we explore a way of diminishing the grain boundary contribution to the spin relaxation by controlling the growth conditions of Ag. For this purpose, we have fabricated Ni$_{80}$Fe$_{20}$ (permalloy, Py)/Ag LSVs using an alternative fabrication process where the Ag channel is epitaxially grown. The epitaxial growth ensures that Ag grains will be well aligned, reducing the grain misalignment and enhancing the transport phenomena. From non-local measurements we determine the spin transport properties of this epitaxial Ag channel, which are superior to those from polycrystalline Ag, which we also prepared as a reference channel structure.

The fabrication of LSVs involves two metallization processes, one for the FM and the other for the NM metal. There are two common techniques for the fabrication process (i) a two-step electron-beam lithography (eBL) followed by metal deposition and lift off [4,6,16] and (ii) a two-angle shadow evaporation technique, where a single eBL step is required [8,9,13]. The only difference between them is that the two-step eBL process needs an extra milling step to obtain a clean FM/NM interface. In both approaches, FM electrodes are usually deposited first and in the second step the NM material is deposited into the channel. However, if epitaxial Ag is grown the standard fabrication process needs to be changed in order to grow the Ag first and the FM metals afterwards.

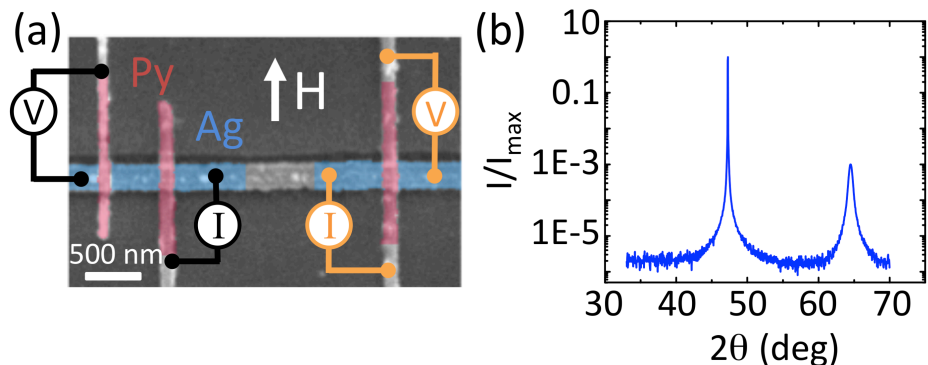

Figure 1. (a) Colored scanning electron microscope (SEM) image of a Py/Ag LSV and a Py bar on top of Ag, used to measure the interface resistance of Py/Ag. The FM and NM materials, the applied magnetic field ($H$) direction and the non-local (black) and interface resistance (orange) measurement configurations are schematically depicted. (b) X-ray diffraction spectra, where the characteristic peak of Si (220) appears at $2\theta \simeq 47.30°$ and the peak of epitaxial Ag (220) appears at $2\theta \simeq 64.45°$.

Thin films with 40 nm of epitaxial Ag were grown at room temperature by sputtering on a (110) Si substrate, after first removing the native Si-oxide by etching the Si-substrate with hydrofluoric (HF) acid [29,30]. The epitaxial growth was confirmed by coplanar $\theta$-$2\theta$ X-Ray diffraction measurements (Fig. 1(b)), where the (220) diffraction peak at $2\theta \simeq 64.45°$ corresponds to the Ag (220) atomic planes. After Ag deposition, the sample was coated with negative resist and, in an initial eBL step, a ~200-nm-wide channel was patterned. Ag was removed with two consecutive Ar-ion etchings (Fig. 2(a)). In the first etching, Ar ions were accelerated almost perpendicularly (80º from in-plane orientation) to the Ag surface in order to remove the Ag that was not protected by the negative resist. In this first step, some etched Ag was redeposited at the edges of the channel, forming vertical walls of Ag that needed to be removed. Therefore, a second etching was performed without breaking the vacuum by accelerating Ar ions almost perpendicular to these Ag walls (10º from in-plane orientation). The suppression of the redeposited metal was confirmed by observing cross-sectional cuts, produced by means of focused ion beam (FIB) irradiation after the first (Fig. 2(b)) and the second etching (Fig. 2(c)). After these etching processes, the FM electrodes were patterned in a second eBL step, using a positive resist in this case. 45-nm-thick Py was e-beam evaporated at a base pressure of ≤ $1\times10^{-8}$ mbar. Different Py electrode widths, ~110 nm and ~150 nm, were chosen in order to obtain different magnetic switching fields. Each sample contains several LSVs where the edge-to-edge distance $L$ between the Py electrodes varied between 150 and 5500 nm. For comparison, a control sample was fabricated following the same process, except that Ag was deposited without pretreating the Si substrate with HF acid, thus leaving the native oxide and leading to a polycrystalline Ag channel structure [31].

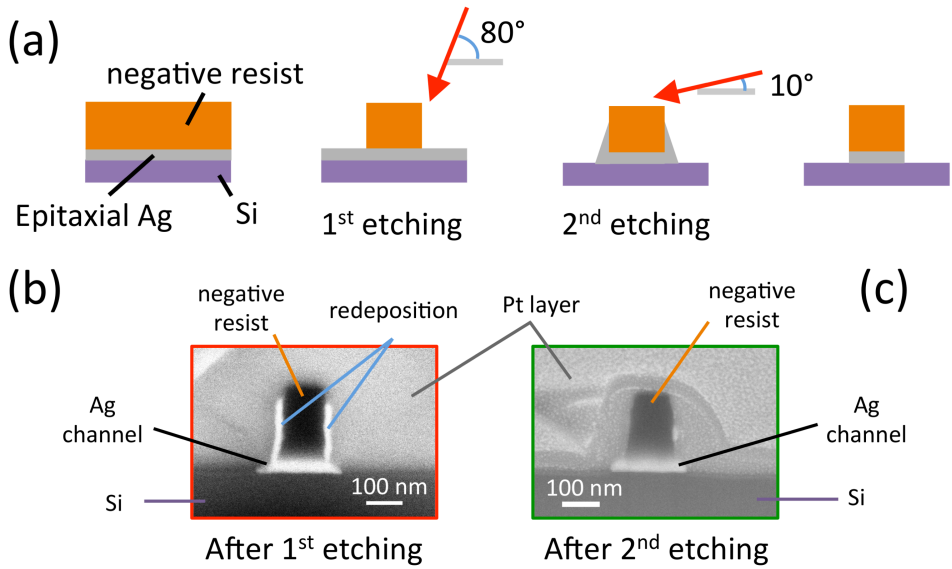

Figure 2. (a) Schematic representation of the two-step Ag milling. (b) Cross-sectional SEM image of the Ag channel just after the first etching. The vertical walls are redeposited Ag. (c) Cross-sectional SEM image of the Ag channel after the second etching. Vertical walls have been milled and the Ag channel has acquired the desired shape. Before cutting the cross sections by FIB, an initial e-beam induced deposition of Pt followed by ion-beam induced deposition of Pt was placed on top of Ag to protect the nanostructure, this is evident in the SEM images.

All measurements described in the following were carried out in a liquid-He cryostat (applying an external magnetic field $H$ and varying the temperature $T$) using a "DC reversal" technique [9]. When a spin-polarized charge current is injected through the Py electrode, due to the net spin polarization of FM materials, a spin accumulation will be created at the Py/Ag interface and will diffuse to both sides of the Ag channel. The second Py electrode will detect the spin accumulation by measuring the voltage between the Py detector and the Ag channel. The measured voltage, $V$, normalized to the injected current, $I$, is defined as the non-local resistance $R_{NL} = \frac{V}{I}$ (See Fig. 1(a) for the measurement scheme). $R_{NL}$ changes sign from positive to negative when the magnetization of the electrodes switches from parallel (P) to antiparallel (AP). We will call this change in resistance the spin signal $\Delta R_{NL}$ (Fig. 3(a)). As $\Delta R_{NL}$ is proportional to the spin accumulation at the detector, $\Delta R_{NL}$ will decay upon varying the distance $L$ at which the spin signal is detected (Fig 3(b)). Solving the one-dimensional spin-diffusion equation to our geometry, the following expression is obtained for $\Delta R_{NL}$ [15,32]:

$$\Delta R_{NL} = \frac{4 R_{Ag}\left[\alpha_i\left(\frac{R_i}{R_{Ag}}\right)+\alpha_{Py}\left(\frac{R_{Py}}{R_{Ag}}\right)\right]^2 e^{-L/\lambda_{Ag}}}{\left[1+2\left(\frac{R_i}{R_{Ag}}\right)+2\left(\frac{R_{Py}}{R_{Ag}}\right)\right]^2 - e^{-2L/\lambda_{Ag}}} \quad (1)$$

where $R_i$ is the interface resistance, $R_{Ag} = \lambda_{Ag}\rho_{Ag}/w_{Ag} t_{Ag}$ and $R_{Py} = \lambda_{Py}\rho_{Py}/(1-\alpha_{Py}^2)w_{Ag} w_{Py}$ are the spin resistances, *i.e.* the tendency of the metals to absorb spin currents, $\lambda_{Ag,Py}$ are the spin diffusion lengths, $\rho_{Ag,Py}$ are the resistivities and $w_{Ag,Py}$ and $t_{Ag}$ are the geometrical parameters (width and thickness) of Ag and Py, respectively. $\alpha_{Py}$ and $\alpha_i$ are the spin polarizations of the Py and the interface, respectively.

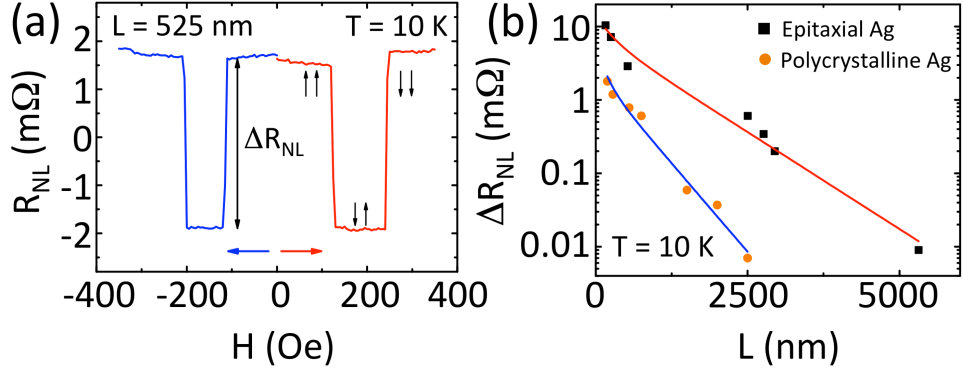

Figure 3. (a) Non-local resistance as a function of the applied magnetic field *H* at 10 K for a Py/Ag LSV with epitaxial Ag where *L*=525 nm. The spin signal is tagged as $\Delta R_{NL}$. (b) Spin signal as a function of the distance *L* between the electrodes at 10 K in Py/Ag LSVs. The solid black (orange) squares (circles) are the experimental data for epitaxial (polycrystalline) Ag and the red (blue) solid line is the fit to Eq. (1).

A $R_i$ of 60 m$\Omega$ is measured in the same device in which the spin signal is obtained (Fig. 1(a)). This measured value is in agreement with a non-transparent interface present in Py/Ag as previously observed [16,20,22]. The origin of this interface resistance is due to the formation of a natural oxide in the Py [20]. By measuring the resistance for every *L* and performing a linear regression, $\rho_{Ag}$ (= 1.06 µ$\Omega$ cm) is obtained, whereas $\rho_{Py}$ (= 22.4 µ$\Omega$ cm) is measured separately, in a device for which Py was grown under the same evaporation conditions. By setting $\lambda_{Py}$= 5 nm [33] and $\alpha_{Py}$= 0.33 [6,7,9,34] we fit our experimental data to Eq. (1) and we obtain the fitting parameters $\alpha_i = 0.47 \pm 0.04$ and $\lambda_{Ag} = 823 \pm 59$ nm at 10 K for epitaxially grown Ag. For comparison, the control sample with polycrystalline growth yields a higher Ag resistivity, $\rho_{Ag} = 2.22$ µ$\Omega$ cm, and a lower spin diffusion length, $\lambda_{Ag} = 449 \pm 30$ nm at 10 K, comparable to other polycrystalline Ag samples reported in literature [17,18].

This substantial improvement in the spin diffusion length, by a factor of two, can be related to the decrease of the spin relaxation via grain boundary scattering [7,15]. For the polycrystalline Ag case as the grains do not have a unique preferred crystallographic orientation, the existing grain boundaries are high angle grain boundaries, which contribute to a higher resistivity, $\rho_{Ag} \backsim 2.22$ μΩ cm, and therefore a shorter $\lambda_{Ag}$. Careful epitaxial growth of Ag strongly reduces the grain boundaries in the channel, lowering the resistivity down to $\rho_{Ag} \backsim 1.07$ μΩ cm and increasing $\lambda_{Ag}$. This dependence is in good agreement with the EY mechanism, which predicts $\lambda_{Ag} \propto \frac{1}{\rho_{Ag}}$. This mechanism is probably similar to what a thermal annealing might do to polycrystalline Ag. For LSVs where Ag has not been treated, $\lambda_{Ag}$~550 nm [17,18] is obtained, whereas values of $\lambda_{Ag}$~1000 nm have been reported after thermally treating the devices [15,21].

In conclusion, we have shown that the spin diffusion length in Ag can be substantially increased by controlling the growth process. When epitaxial Ag is grown, the grain boundary scattering is largely suppressed leading to lower resistivity values and higher spin diffusion lengths. The main advantage that this approach offers compared to an annealing treatment is that the growth process is done at room temperature. This avoids a possible diffusion of metals when the device is being heated. Proper engineering of the material used as a spin channel can thus improve the spin transport properties, and hereby help towards the development of devices based on pure spin currents.

**Acknowledgments**

This work is supported by the European Union 7th Framework Program under the Marie Curie Actions (256470-ITAMOSCINOM) and the European Research Council (257654-SPINTROS), by the Spanish MINECO (MAT2012-37638 and MAT2012-36844) and by the Basque Government (PI2011-1 and PI2012-47). M. I., E. V., L. F. and O. I. thank the Basque Government for a PhD fellowship (BFI-2011-106, BFI-2010-163, PRE-2013-1-974 and BFI-2009-284, respectively).